\documentclass[prd,aps,nofootinbib,floatfix,10pt]{revtex4}

\usepackage{amsmath,graphicx,epsfig,amssymb,dsfont,mathtools}
\usepackage[usenames]{color}
\usepackage{ulem} 
\usepackage{bigstrut}
\usepackage{slashed}
\usepackage{multirow}

\begin{document}
\title{Study of $s\to d\nu\bar{\nu}$ rare hyperon decays in the Standard Model and new physics}
\author{Xiao-Hui Hu$^1$~\footnote{Email:huxiaohui@sjtu.edu.cn}, Zhen-Xing Zhao$^1$~\footnote{Email:star\_0027@sjtu.edu.cn}}
\affiliation{$^{1}$ INPAC, Shanghai Key Laboratory for Particle Physics and Cosmology,\\
                MOE Key Laboratory for Particle Physics, Astrophysics and Cosmology,\\
                School of Physics and Astronomy, Shanghai Jiao-Tong University, Shanghai
                200240, P.R. China }

\begin{abstract}
	FCNC processes offer important tools to test the Standard Model (SM), and to search
	for possible new physics. In this work, we investigate the  $s\to d\nu\bar{\nu}$ rare
	hyperon decays in SM and beyond. We find that in SM the branching
	ratios for  these rare hyperon decays range from $10^{-14}$ to $10^{-11}$. When all the errors in the form factors are included, we find that the final branching fractions for most decay modes have an uncertainty of about $5\%$ to $10\%$.
	After taking into account the contribution from new physics, the generalized SUSY extension of SM and the minimal 331 model,
	the decay widths for these channels can be enhanced by a factor of $2 \sim 7$.
\end{abstract}

\maketitle
	
\section{Introduction}

The flavor changing neutral current (FCNC) transitions
provide
a critical test of the Cabibbo-Kobayashi-Maskawa (CKM) mechanism in
the Standard Model (SM), and allow to search for the possible new physics. In SM, FCNC transition $s\to d\nu\bar{\nu}$ proceeds through
$Z$-penguin and electroweak box diagrams, and thus the decay probablitities are strongly
suppressed. In
this case, a precise study allows to perform very stringent tests
of SM and ensures large sensitivity to potential new degrees
of freedom.

A large number of studies have been performed of the $K^{+}\to\pi^{+}\nu\bar{\nu}$
and $K_{L}\to\pi^{0}\nu\bar{\nu}$ processes, and reviews of these two decays can
be found in \cite{Buras:2013ooa,Buras:2004uu,Komatsubara:2012pn,Blanke:2013goa,Smith:2014mla,Buras:2015hna}.
On the theoretical side, using the most
recent input parameters, the SM predictions
for the two branching ratios are \cite{Buras:2015qea}
\begin{align}
{\cal B}(K^{+}\to\pi^{+}\nu\bar{\nu})_{{\rm SM}} & =(8.4\pm1.0)\times10^{-11},\label{eq:SMk+pi+}\\
{\cal B}(K_{L}\to\pi^{0}\nu\bar{\nu})_{{\rm SM}} & =(3.4\pm0.6)\times10^{-11}.
\end{align}
The dominant uncertainty comes from the CKM
matrix elements and the charm contribution. On the experimental
side, the NA62 experiment at the CERN SPS has reported the first search for $K^{+}\to\pi^{+}\nu\bar{\nu}$ using the decay-in-flight technique, and
the corresponding observed upper limit is \cite{CortinaGil:2018fkc}:
\begin{equation}
\label{exp:kpinunu}
{\cal B}(K^{+}\to\pi^{+}\nu\bar{\nu})_{{\rm exp}}<14\times 10^{-10},\quad \rm{at\, 95\%CL}.
\end{equation}
Similarly, the E391a collaboration reported the 90\% C.L. upper bound \cite{Ahn:2009gb}
\begin{equation}\label{exp:klpinunu}
{\cal B}(K_{L}\to\pi^{0}\nu\bar{\nu})_{{\rm exp}}\le2.6\times10^{-8}.
\end{equation}

The KOTO experiment, an upgrade of the E391a experiment,
aims at a first observation of the $K_{L}\to\pi^{0}\nu\bar{\nu}$
decay at J-PARC around 2020 \cite{Komatsubara:2012pn,Shiomi:2014sfa}. Given the goal of a 10\% precision by NA62, the authors
of Ref. \cite{Christ:2016lro} intend to carry out lattice QCD calculations
to determine the long-distance contributions to the $K^{+}\to\pi^{+}\nu\bar{\nu}$
amplitude.

Analogous to $K^{+}\to\pi^{+}\nu\bar{\nu}$ and $K_{L}\to\pi^{0}\nu\bar{\nu}$,
the rare hyperon decays $B_{i}\to B_{f}\nu\bar{\nu}$ also proceed
via $s\to d\nu\bar{\nu}$ at the quark level, and thus offer important tools
to test SM, and to search for possible new physics. Compared
to the widely considered $K^{+}\to\pi^{+}\nu\bar{\nu}$ and $K_{L}\to\pi^{0}\nu\bar{\nu}$,
there are few studies devoted to the rare hyperon decays $B_{i}\to B_{f}\nu\bar{\nu}$.
This work aims to perform a preliminary theoretical research of the rare hyperon
decays both in and beyond SM.

A study of the hyperon decays at the BESIII experiment is proposed using the hyperon parents from $J/\psi$ decay. The electron-positron collider BEPCII provides a clean experimental environment. About
$10^{6}$-$10^{8}$ hyperons, $\Lambda$, $\Sigma$, $\Xi$ and $\Omega$,
are produced in the $J/\psi$ and $\psi(2S)$ decays with the
proposed data samples at the BESIII experiment. Based on these samples,
the sensitivity for the measurement of the branching ratios of the
hyperon decays is in the range of $10^{-5}$-$10^{-8}$.  The author
of Ref. \cite{Li:2016tlt} proposed that rare decays and decays with
invisible final states may be probed.

The paper is organized as follows. In Sec. II, our computing
framework is presented. Sec. III is devoted to performing the
numerical calculations. The branching ratios of several rare hyperon
decays are calculated in SM. The new physic contribution, the Minimal Supersymmetric Standard Model (MSSM) and the minimal 331 model, are considered. We also discuss possible uncertainties from the form factors.
The last section contains a short summary.

\section{Theoretical framework}

The next-to-leading order (NLO) effective Hamiltonian for $s\to d\nu\bar{\nu}$ reads \cite{Buchalla:1995vs}:
\begin{equation}
{\cal H}_{{\rm eff}}=\frac{G_{F}}{\sqrt{2}}\frac{\alpha}{2\pi\sin^{2}\theta_{W}}\sum_{l=e,\mu,\tau}[V_{cs}^{*}V_{cd}X_{N{\rm L}}^{l}+V_{ts}^{*}V_{td}X(x_{t})](\bar{s}d)_{V-A}(\bar{\nu}_{l}\nu_{l})_{V-A}+{\rm h.c.},\label{eq:Heff_SM}
\end{equation}
where $X(x_{t})$ and $X_{N{\rm L}}^{l}$ are relevant for the top
and the charm contribution, respectively. Their explicit expressions
can be found in Ref. \cite{Buchalla:1995vs}.
Here, $x_t=m_t^2/m^2_W$. To leading order in $\alpha_s$, the function $X(x_t)$ relevant for the top contribution reads \cite{Inami:1980fz,Buchalla:1993bv}
\begin{align}
\label{xx}
X(x)&=X_0(x)+{\alpha_s\over4\pi} X_1(x), \nonumber\\
X_0(x)&={x\over 8}\left[ -{2+x\over 1-x}+{3x-6\over (1-x)^2}\ln x\right],
\nonumber\\
X_1(x)&=-{23x+5x^2-4x^3\over 3(1-x)^2}+{x-11x^2+x^3+x^4\over (1-x)^3}\ln x
+{8x+4x^2+x^3-x^4\over 2(1-x)^3}\ln^2 x \nonumber\\
&\quad-{4x-x^3\over (1-x)^2}L_2(1-x)+8x{\partial X_0(x)\over\partial x}\ln x_\mu,
\end{align}
where $x_\mu=\mu^2/M^2_W$ with $\mu={\cal O}(m_t)$ and
\begin{equation}
\label{l2} L_2(1-x)=\int^x_1 dt {\ln t\over 1-t}.
\end{equation}
The function $X^l_{NL}$ corresponds to $X(x_t)$ in the charm sector. It results from the renormalization group (RG) calculation in next-to-leading-order logarithmic approximation (NLLA) and is given as follows:
\begin{equation}
\label{xlnl}X^l_{NL}=C_{NL}-4 B^{(1/2)}_{NL},
\end{equation}
where $C_{NL}$ and $B^{(1/2)}_{NL}$ correspond to the $Z^0$-penguin and the
box-type contribution, respectively, given as \cite{Buchalla:1993wq}
\begin{align}
\label{cnln}
C_{NL}
&={x(m_c)\over 32}K^{{24\over 25}}_c\left[\left({48\over 7}K_++
 {24\over 11}K_--{696\over 77}K_{33}\right)\left({4\pi\over\alpha_s(\mu)}+
 {15212\over 1875} (1-K^{-1}_c)\right)\right.\nonumber\\
&\quad+\left(1-\ln{\mu^2\over m_c^2}\right)(16K_+-8K_-)-{1176244\over 13125}K_+-
 {2302\over 6875}K_-+{3529184\over 48125}K_{33} \nonumber\\
&\quad+\left. K\left({56248\over 4375}K_+-{81448\over 6875}K_-+{4563698\over 144375}K_{33} \right)\right],\nonumber\\
B^{(1/2)}_{NL}&={x(m_c)\over 4}K^{24\over 25}_c\left[ 3(1-K_2)\left(
 {4\pi\over\alpha_s(\mu)}+{15212\over 1875}(1-K^{-1}_c)\right)\right.\nonumber\\
&\quad-\left.\ln{\mu^2\over m_c^2}-
  {r\ln r\over 1-r}-{305\over 12}+{15212\over 625}K_2+{15581\over 7500}K K_2
  \right],
\end{align}
where $r=m^2_l/m^2_c(\mu)$, $\mu={\cal O}(m_c)$ and
\begin{equation}
\label{kkc}
K={\alpha_s(M_W)\over\alpha_s(\mu)},\qquad
  K_c={\alpha_s(\mu)\over\alpha_s(m_c)} ,\qquad
 K_+=K^{{6\over 25}},\qquad
 K_-=K^{-{12\over 25}},\qquad
 K_{33}=K_2=K^{-{1\over 25}}.
 \end{equation}

In the following, we consider the transitions between the baryon octet ($\Xi$, $\Sigma$, $\Lambda$ and $N$) and the transitions from the baryon decuplet to the octet $\Omega^-\to\Xi^-$.

The transition matrix elements of the vector and axial-vector currents between the baryon octets can be parametrized in terms of six form factors $f_{1,2,3}(q^2)$ and $g_{1,2,3}(q^2)$:
\begin{align}
\langle B^{\prime}_8(P^{\prime},S_{z}^{\prime})|\bar{d}\gamma_{\mu}(1-\gamma_{5})s|B_8(P,S_{z})\rangle & =\bar{u}(P^{\prime},S_{z}^{\prime})\left[\gamma_{\mu}f_{1}(q^{2})+i\sigma_{\mu\nu}\frac{q^{\nu}}{M}f_{2}(q^{2})+\frac{q_{\mu}}{M}f_{3}(q^{2})\right]u(P,S_{z})\nonumber \\
& -\bar{u}(P^{\prime},S_{z}^{\prime})\left[\gamma_{\mu}g_{1}(q^{2})+i\sigma_{\mu\nu}\frac{q^{\nu}}{M}g_{2}(q^{2})+\frac{q_{\mu}}{M}g_{3}(q^{2})\right]\gamma_{5}u(P,S_{z}),\label{eq:Lorentz}
\end{align}
where $q=P-P^{\prime}$, and $M$ denotes the mass of the parent baryon octet $B_8$.
The form factors for $B_8\to B_8^{\prime}$ transition $f_i(q^2)$ and $g_i(q^2)$ can be expressed by the following formulas~\cite{Gaillard:1984ny}:
\begin{align}
f_{m}=aF_m(q^2)+bD_m(q^2), \quad
g_{m}=aF_{m+3}(q^2)+bD_{m+3}(q^2), \quad (m=1,2,3), \label{eq:formfactor}
\end{align}
where $F_i(q^2)$ and $D_i(q^2)$ with $i=1,2,\cdots,6$ are different functions of $q^2$ for each of the six form factors. Some remarks are necessary~\cite{Gaillard:1984ny}:
\begin{itemize}
\item The constants $a$ and $b$ in Eq.~(\ref{eq:formfactor}) are SU(3) Clebsch-Gordan coefficients that appear when an octet operator is sandwiched between octet states.
\item For $q^2 = 0$, the form factor $f_1(0)$ is equal to the electric charge of the baryon, therefore $F_1(0)=1$ and $D_1(0)=0$.
\item The weak $f_2(0)$ form factor can be computed using the anomalous magnetic moments of proton and neutron ($\kappa_p$ and $\kappa_n$) in the exact SU(3) symmetry. Here $F_2(0) =\kappa_p + \frac{1}{2}\kappa_n$ and $D_2(0) = -\frac{3}{2}\kappa_n$.
\item $g_1(0)$ is a linear combination of two parameters, $F$ and $D$.
\item  Since $g_2^{n\to p}=F_5(q^2)+D_5(q^2)=0$ and $ g_2^{\Xi^{-}\to\Xi^{0}}=D_5(q^2)-F_5(q^2)=0$, we could get $F_5(q^2)=D_5(q^2)=0$. Therefore, all the pseudo-tensor form factors $g_2$ vanish in all decays up to symmetry-breaking effects.
\item In the $s\to d\bar\nu\nu$ decay, the $f_3$ and $g_3$ terms are proportional to the neutrino mass and thus can be neglected for the decays considered in this work.
\end{itemize}

Since the invariant mass squared of lepton pairs in the hyperon decays is relatively small, it is expected that the $q^2$ distribution in the form factors has small impact on the decay widths. We list the expressions for $f_1$, $f_2$ and $g_1$ at $q^2=0$ in TAB.~\ref{Tab:formfactors}.

\begin{table}
\caption{The form factors for $B\to B'$ transition $f_1(0)$, $f_2(0)$ and $g_1(0)$~\cite{Gaillard:1984ny}, where the experimental anomalous magnetic moments are $\kappa_p=1.793\pm0.087$ and $\kappa_n=-1.913\pm0.069$~\cite{Yang:2015era}, with the two coupling constants $F=0.463\pm0.008$ and $D=0.804\pm0.008$~\cite{Yang:2015era}. Here, $g_1/f_1$ is positive for the neutron decay, and all other signs are fixed using this sign convention.}
	\label{Tab:formfactors}
	\begin{center}
		\begin{tabular}{c|c|c|c|c|c}
			\hline\hline
  			$B\to B'$ & $\Lambda\to n$ & $\Sigma^{+}\to p$
			& $\Xi^0\to\Lambda$&$\Xi^0\to\Sigma^0$
			&$\Xi^-\to\Sigma^-$\tabularnewline
			\hline
			$f_1(0)$ & $-\sqrt{\frac{3}{2}}$ & $-1$&$\sqrt{\frac{3}{2}}$&-$\frac{1}{\sqrt{2}}$&1\tabularnewline
			\hline
			$f_2(0)$  &$ -\sqrt{\frac{3}{2}}\kappa_{p}$ & $-(\kappa_p+2\kappa_n) $ & $\sqrt{\frac{3}{2}}(\kappa_p+\kappa_n) $&$-\frac{1}{\sqrt{2}}(\kappa_p-\kappa_n)$&$\kappa_p-\kappa_n$\tabularnewline
			\hline
			$g_1(0)$ & $ -\sqrt{\frac{3}{2}}(F+D/3)$& $-(F-D) $ & $\sqrt{\frac{3}{2}}(F-D/3) $&$-\frac{1}{\sqrt{2}}(F+D) $&$F+D$\tabularnewline
			\hline\hline
		\end{tabular}
	\end{center}
\end{table}
Hence, Eq. (\ref{eq:Lorentz}) can be rewritten as:
\begin{align}
\langle B^{\prime}_8(P^{\prime},S_{z}^{\prime})|\bar{d}\gamma_{\mu}(1-\gamma_{5})s|B_8(P,S_{z})\rangle & =\bar{u}(P^{\prime},S_{z}^{\prime})\left[\gamma_{\mu}f_{1}(q^{2})+i\sigma_{\mu\nu}\frac{q^{\nu}}{M}f_{2}(q^{2})-\gamma_{\mu}g_{1}(q^{2})\gamma_{5}\right]u(P,S_{z}).
\label{eq:Lorentz1}
\end{align}
The helicity amplitudes of hadronic part are defined as
\begin{align}
&H_{\lambda^{\prime},\lambda_{V}}^{V}\equiv\langle B^{\prime}_8(P^{\prime},\lambda^{\prime})|\bar{d}\gamma^{\mu}s|B_8(P,\lambda)\rangle\epsilon_{V\mu}^{*}(\lambda_{V}),\label{eq:heli-amp-V}\\
&H_{\lambda^{\prime},\lambda_{V}}^{A}\equiv\langle B^{\prime}_8(P^{\prime},\lambda^{\prime})|\bar{d}\gamma^{\mu}\gamma_{5}s|B_8(P,\lambda)\rangle\epsilon_{V\mu}^{*}(\lambda_{V}).\label{eq:heli-amp-A}
\end{align}
Here $\lambda^{(\prime)}$ denotes the helicity of the parent (daughter)
baryon in the initial (final) state, and $\lambda_{V}$ is the helicity
of the virtual intermediate vector particle. It can be shown that
the helicity amplitudes $H_{\lambda^{\prime},\lambda_{V}}^{V,A}$ have
the following simple forms \cite{Ke:2007tg}:
\begin{align}
H_{\frac{1}{2},0}^{V} & =-i\frac{\sqrt{Q_{-}}}{\sqrt{q^{2}}}\left[(M+M^{\prime})f_1-\frac{q^2}{M}f_2\right],\quad H_{\frac{1}{2},0}^{A}  =-i\frac{\sqrt{Q_{+}}}{\sqrt{q^{2}}}(M-M^{\prime})g_1,\nonumber  \\
H_{\frac{1}{2},1}^{V} & =i\sqrt{2Q_{-}}\left[-f_1+\frac{M+M^{\prime}}{M}f_2\right],\,\,\,\qquad
 H_{\frac{1}{2},1}^{A}  =-i\sqrt{2Q_{+}}g_1.\label{eq:heli-amp-con}
\end{align}

In the above, $Q_{\pm}=(M\pm M^{\prime})^{2}-q^{2}$, and $M$ ($M^{\prime}$)
is the parent (daughter) baryon mass in the initial (final) state.
The amplitudes for the negative helicity  are obtained from
the relations,
\begin{equation}
H_{-\lambda^{\prime},-\lambda_{V}}^{V}=H_{\lambda^{\prime},\lambda_{V}}^{V},\qquad \qquad H_{-\lambda^{\prime},-\lambda_{V}}^{A}=-H_{\lambda^{\prime},\lambda_{V}}^{A}.
\end{equation}
The complete helicity amplitudes are obtained by
\begin{equation}
H_{\lambda^{\prime},\lambda_{V}}=H_{\lambda^{\prime},\lambda_{V}}^{V}-H_{\lambda^{\prime},\lambda_{V}}^{A}.
\end{equation}

Due to the  lack of experimental data for the $M_1$ and $E_2$ transitions from the baryon decuplet to the octet, the vector transition matrix element for $\Omega^-\to \Xi^-$ can not be determined. In this work we follow Ref.~\cite{Yang:2015era}. and consider only the  axial-vector current  matrix element~\cite{Yang:2015era,Alexandrou:2006mc,LlewellynSmith:1971uhs}:
\begin{align}
\langle \Xi^-(P',S'_z)|\bar{d}\gamma_{\mu}\gamma_5s|\Omega^-(P,S_z)\rangle&=
\bar{u}_{\Xi^-}(P',S'_z)\left\{C_5^A(q^2)g_{\mu\nu}+C_6^{A}(q^2)q_{\mu}q_{\nu}\right.\nonumber\\
&\quad\quad+\left.\left[C_{3}^{A}(q^2)\gamma^{\alpha}+
C_4^{A}(q^2)p'\right](q_{\alpha}g_{\mu\nu}-q_{\nu}g_{\alpha\mu})\right\}
u^{\nu}_{\Omega^-}(P,S_z).
\end{align}
Here, $u^{\nu}_{\Omega^-}(P,S_z)$ represents the Rarita-Schwinger spinor that describes the baryon decuplet $\Omega^-$ with spin $\frac{3}{2}$. In Ref.~\cite{Berman:1965iu} it is shown that $C_3^A(q^2)$ and $C_4^A(q^2)$ are proportional to the mass difference of the initial and final baryons, and thus are suppressed. In the chiral limit, $C_5^A(q^2)$ and $C_6^A(q^2)$ are related by $C_6^A(q^2)=M_{N}^2C_5^A(q^2)/q^2$~\cite{Alexandrou:2006mc}. In our calculations, we use $C_5^A(0)=1.653\pm0.006$ for $\Omega^-\to\Xi^-$,  which is the same as $\Omega^-\to\Xi^0$ in the SU(3) limit~\cite{Yang:2015era}.
The helicity amplitude can be expressed as:
\begin{align}
H^A_{\lambda',\lambda_{V}}&=\langle \Xi^-(P',\lambda')|\bar{d}\gamma_{\mu}\gamma_5s|\Omega^-(P,\lambda)\rangle\epsilon^{*\mu}_{V}(\lambda_{V})\\
&=\bar{u}_{\Xi^{-}}(P',\lambda')
\left[C_{5}^A(q^2)g_{\mu\nu}+
C_6^A(q^2)q_{\mu}q_{\nu}\right]
u_{\Omega^-}^\nu(P,\lambda)\epsilon^{*\mu}_{V}(\lambda_{V}).
\end{align}
Here, $\lambda^{(\prime)}$ and $\lambda_{V}$ have the same definition as in Eqs.(\ref{eq:heli-amp-V})-(\ref{eq:heli-amp-A}). It can be shown that the helicity amplitudes $H_{\lambda^{\prime},\lambda_{V}}^{A}$ have
the following simple forms \cite{Ke:2007tg}:
\begin{align}
H^A_{\frac{1}{2},0}=H^A_{-\frac{1}{2},0}=i\sqrt{\frac{2Q_{+}}{3}}\frac{E_{V}}{\sqrt{q^2}}C_{5}^A(q^2),\quad
H^A_{\frac{1}{2},1}=H^A_{-\frac{1}{2},-1}=i\sqrt{\frac{Q_{+}}{3}}C_{5}^A(q^2),\quad
H^A_{\frac{1}{2},-1}=H^A_{-\frac{1}{2},1}=i\sqrt{Q_{+}}C_{5}^A(q^2).
\end{align}
The differential decay width for $B\to B^{\prime}\bar\nu\nu$ is given as:
\begin{equation}
\frac{d\Gamma}{dq^{2}}=\frac{d\Gamma_{L}}{dq^{2}}+\frac{d\Gamma_{T}}{dq^{2}}.
\end{equation}
Here, $d\Gamma_{L}/dq^{2}$ and $d\Gamma_{T}/dq^{2}$
are the longitudinal and transverse parts of the decay width, and their explicit expressions are given by
\begin{align}
\frac{d\Gamma_{L}}{dq^{2}}&=N\frac{q^{2}p^{\prime}}{12(2\pi)^{3}M^{2}}(|H_{\frac{1}{2},0}|^{2}+|H_{-\frac{1}{2},0}|^{2}),\label{eq:longi}\\
\frac{d\Gamma_{T}}{dq^{2}}&=N\frac{q^{2}p^{\prime}}{12(2\pi)^{3}M^{2}}(|H_{\frac{1}{2},1}|^{2}+|H_{-\frac{1}{2},-1}|^{2}+|H_{\frac{1}{2},-1}|^{2}+|H_{-\frac{1}{2},1}|^{2}).\label{eq:trans}
\end{align}
In Eqs.~(\ref{eq:longi}) and (\ref{eq:trans}), $p^{\prime}=\sqrt{Q_{+}Q_{-}}/2M$ is the magnitude
of the momentum of $B^{\prime}$ in the rest frame of $B$, and $N=2N_{1}(0)+N_{1}(m_{\tau})$
with
\begin{equation}
N_{1}(m_{l})=\left|\frac{G_{F}}{\sqrt{2}}\frac{\alpha}{2\pi\sin^{2}\Theta_{W}}\left(V_{cd}^{*}V_{cs}X_{NL}^{l}(m_{l})+V_{td}^{*}V_{ts}X(x_{t})\right)\right|^{2}.
\end{equation}
Note that we have neglected the electron and muon masses.

One can then obtain the decay width
\begin{equation}
\Gamma=\int_{0}^{(M-M^{\prime})^{2}}dq^{2}\frac{d\Gamma}{dq^{2}}.
\end{equation}

\section{Numerical results and discussion}

\subsection{Calculations in SM}

\begin{table}
\caption{The input parameters of this work.}
\label{Tab:inputsparameters}
\begin{center}
\begin{tabular}{l|l|l|l|l}
\hline
\hline
\multicolumn{5}{c}{The masses and lifetimes of baryons in the initial and final states\cite{Tanabashi:2018oca}}\tabularnewline
\hline
$m_{p}=938.2720813$MeV&
\multicolumn{2}{l|}{$m_{\Sigma^{+}}=1189.37$MeV}&
\multicolumn{2}{l}{$m_{\Xi^{0}}=1314.86$MeV}
\tabularnewline\hline
$m_{n}$$=939.5654133$MeV&
 \multicolumn{2}{l|}{$m_{\Sigma^{-}}=1197.45$MeV} &
 \multicolumn{2}{l}{$m_{\Xi^{-}}=1321.71$MeV}\tabularnewline
\hline
$m_{\Lambda}=1115.683$MeV&
\multicolumn{2}{l|}{$m_{\Sigma^{0}}=1192.642$MeV}&
\multicolumn{2}{l}{$m_{\Omega^{-}}=1672.45$MeV}
\tabularnewline
\hline
$\tau_{\Xi^{0}}=2.90\times10^{-10}s$ & \multicolumn{2}{l|}{$\tau_{\Xi^{-}}=1.639\times10^{-10}s$}&
\multicolumn{2}{l}{$\tau_{\Omega^{-}}=0.821\times10^{-10}s$}
\tabularnewline
\hline
$\tau_{\Lambda}=2.632\times10^{-10}s$ & \multicolumn{2}{l|}{$\tau_{\Sigma^{+}}=0.8018\times10^{-10}s$}&
\multicolumn{2}{r}{}\tabularnewline
\hline
\multicolumn{5}{c}{Physical constants and CKM Parameters\cite{Tanabashi:2018oca,Buras:2005gr}}\tabularnewline
\hline
$G_F=1.16637387\times10^{-5}{\rm GeV}^{-2}$ &
$\sin^2\theta_{W}=0.23122$& $\alpha_{s}$($m_{Z}$)$=0.1182$
&\multicolumn{2}{l}{$\alpha\equiv\alpha$($m_{Z}$)$=1/128$}\tabularnewline
\hline
$m_{\ensuremath{\tau}}$=$1776.86$MeV & $m_{c}=1.275$GeV&
$m_{t}=173.0$GeV &$m_{W}$=$80.379$ GeV & $m_{Z}$$=91.1876$GeV\tabularnewline
\hline
$A=0.836$ &
$\lambda=0.22453$&
$\bar{\rho}=0.122$ & \multicolumn{2}{l}{$\bar{\eta}=0.355$}\tabularnewline
\hline
\hline
\end{tabular}
\end{center}
\end{table}
With the inputs parameters given in TAB.~\ref{Tab:inputsparameters} and the formulae from the last section,
the LO and NLO results for $\mu_{c}=1~{\rm GeV},\mu_{t}=100~{\rm GeV}$ and $\mu_{c}=3~{\rm GeV},\mu_{t}=300~{\rm GeV}$
are listed in TAB.~\ref{Tab:branchingfraction}.

From the results in TAB. ~\ref{Tab:branchingfraction} one can see that:
\begin{itemize}
	\item The branching fractions of $s\to d\nu\bar{\nu}$ rare hyperon decays
	range from $10^{-14}$ to $10^{-11}$.
	\item For $\mu_{c}=1{\rm GeV},\mu_{t}=100{\rm GeV}$, the NLO results are
	smaller than the LO ones by about 30\%, while for $\mu_{c}=3{\rm GeV},\mu_{t}=300{\rm GeV}$
	the NLO results are larger than the LO ones by about 10\%.
	\item The LO results vary by about 50\% from $\mu_{c}=1{\rm GeV},\mu_{t}=100{\rm GeV}$
	to $\mu_{c}=3{\rm GeV},\mu_{t}=300{\rm GeV}$, while the NLO ones
	vary by about 30\%. As expected, the NLO results depend less on the
	mass scales.
	\item The branching ratio of $\Omega^{-}\to\Xi^{-}\nu\bar{\nu}$ is the largest among the 6 channels. It is of the same order as for $K^+\to\pi^{+}\nu\bar{\nu}$ and $K_{L}\to\pi^{0}\nu\bar{\nu}$.
\end{itemize}

At present, there is a small number of experimental studies, and thus  most experimental constraints  are less severe. The prospects for rare and forbidden hyperon decays at BESIII were analyzed in a recent publication Ref.\cite{Li:2016tlt}. We quote the experimental sensitivity for all decay modes in TAB.~\ref{Tab:branchingfraction}. Unfortunately, one can see that the current BESIII experiment will not be able to probe these hyperon decays. We hope this may be improved at future experimental facilities like the Super Tau-Charm Factory.
\begin{table}
\caption{The LO, NLO, NLO+SUSY and NLO+$\rm M331$  results for the branching ratio of rare hyperon decays for $\mu_{c}=1~{\rm GeV},\mu_{t}=100~{\rm GeV}$ and $\mu_{c}=3~{\rm GeV},\mu_{t}=300~{\rm GeV}.$}
	\label{Tab:branchingfraction}
\begin{center}
	\begin{tabular}{c|c|c|c|c|c|c|c}
		\hline\hline
		\multicolumn{2}{c|}{Branching ratio} &
		${\cal{B}}(\Lambda\rightarrow n\nu\bar{\nu})$&
		${\cal{B}}(\Sigma^{+}\rightarrow p\nu\bar{\nu})$& ${\cal{B}}(\Xi^{0}\rightarrow\Lambda\nu\bar{\nu})$&
		${\cal{B}}(\Xi^{0}\rightarrow\Sigma^{0}\nu\bar{\nu})$&
		${\cal{B}}(\Xi^{-}\rightarrow\Sigma^{-}\nu\bar{\nu})$& ${\cal{B}}(\Omega^{-}\rightarrow\Xi^{-}\nu\bar{\nu})$\tabularnewline
		\hline
		\multirow{2}{*}{\footnotesize{$\mu_{c}=1$~GeV}}
		&LO &
		$2.85\times10^{-12}$&
		$6.88\times10^{-13}$&
		$1.06\times10^{-12}$&
		$1.77\times10^{-13}$&
		$2.17\times10^{-13}$&
		$1.78\times10^{-11}$\tabularnewline	\cline{2-8}	
		&NLO &
		$1.98\times10^{-12}$&
		$5.01\times10^{-13}$&
		$7.35\times10^{-13}$&
		$1.24\times10^{-13}$&
		$1.52\times10^{-13}$&
		$1.93\times10^{-11}$ \tabularnewline\cline{2-8}
		\multirow{3}{*}{\footnotesize{$\mu_{t}=100$~GeV}}&
		\footnotesize{NLO+SUSY (Set.I)}&
		$8.14\times10^{-12}$&
		$2.06\times10^{-12}$&
		$3.02\times10^{-12}$&
		$5.08\times10^{-13}$&
		$6.23\times10^{-13}$&
		$7.94\times10^{-11}$\tabularnewline\cline{2-8}
		&\footnotesize{NLO+SUSY (Set.II)} &
		$3.78\times10^{-12}$&
		$9.55\times10^{-13}$&
		$1.40\times10^{-12}$&
		$2.36\times10^{-13}$&
		$2.89\times10^{-13}$&
		$3.69\times10^{-11}$ \tabularnewline\cline{2-8}
		& \footnotesize{NLO+$\rm M 331$}&
		$1.24\times10^{-11}$&
		$3.13\times10^{-12}$&
		$4.59\times10^{-12}$&
		$7.71\times10^{-13}$&
		$9.45\times10^{-13}$&
		$1.20\times10^{-10}$ \tabularnewline\hline
		\multirow{2}{*}{\footnotesize$\mu_{c}=3$~GeV}
		&\footnotesize{LO} &
		$1.10\times10^{-12}$&
		$2.65\times10^{-13}$&
		$4.10\times10^{-13}$&
		$6.83\times10^{-14}$&
		$8.37\times10^{-14}$&
		$1.07\times10^{-11}$\tabularnewline	\cline{2-8}	
		&\footnotesize{NLO} &
		$1.20\times10^{-12}$&
		$3.04\times10^{-13}$&
		$4.46\times10^{-13}$&
		$7.50\times10^{-14}$&
		$9.19\times10^{-14}$&
		$1.17\times10^{-11}$ \tabularnewline\cline{2-8}
		\multirow{3}{*}{\footnotesize{$\mu_{t}=300$~GeV}}&
		\footnotesize{NLO+SUSY (Set.I)}&
		$5.85\times10^{-12}$&
		$1.48\times10^{-12}$&
		$2.17\times10^{-12}$&
		$3.65\times10^{-13}$&
		$4.47\times10^{-13}$&
		$5.71\times10^{-11}$ \tabularnewline\cline{2-8}
		&\footnotesize{NLO+SUSY (Set.II)} &
		$2.35\times10^{-12}$&
		$5.94\times10^{-13}$&
		$8.72\times10^{-13}$&
		$1.47\times10^{-13}$&
		$1.80\times10^{-13}$&
		$2.29\times10^{-11}$\tabularnewline\cline{2-8}
		& \footnotesize{NLO+$\rm M 331$}&
		$1.02\times10^{-11}$&
		$2.58\times10^{-12}$&
		$3.80\times10^{-12}$&
		$6.37\times10^{-13}$&
		$7.81\times10^{-13}$&
		$9.95\times10^{-11}$\tabularnewline\hline
\multicolumn{2}{c|}{BESIII sensitivity \cite{Li:2016tlt}}&
		$3\times10^{-7}$&
		$4\times10^{-7}$&
		$8\times10^{-7}$&
		$9\times10^{-7}$&
		$--$&
		$2.6\times10^{-5}$\tabularnewline\hline\hline
	\end{tabular}
\end{center}
\end{table}
\subsection{Form factors Uncertainties}

Note that due to the Ademollo-Gatto theorem~\cite{Ademollo:1964sr}, the form factor $f_1(0)$ does not receive any SU(3) symmetry breaking  correction. However, $f_2(0)$ can be computed using the anomalous magnetic moments of proton and neutron ($\kappa_p$ and $\kappa_n$) in the exact SU(3) symmetry. The experimental data for $\kappa_p$ and $\kappa_n$ already include the  SU(3) symmetry breaking effects~\cite{Yang:2015era}:
\begin{align}
\kappa_p\left[{\cal O}(m^0_s)\right]
&=1.363\pm0.069,\,\,\kappa_n
\left[{\cal O}(m^0_s)\right]=-1.416 \pm 0.049,\\
\kappa_p\left[{\cal O}(m^0_s) + {\cal O}(m^1_s)\right]
&=1.793\pm0.087,\,\,\kappa_n
\left[{\cal O}(m^0_s) + {\cal O}(m^1_s)\right]=-1.913\pm0.069.
\end{align}
The uncertainties from $\kappa_p$ and $\kappa_n$ in the effect of SU(3) symmetry  breaking is approximately $25\%$. We calculated the effect of $\kappa_p$ and $\kappa_n$ on the branching ratio of $\Sigma^+\to p\nu\bar{\nu}$ in the case of NLO with the energy scale $\mu_c=1\rm{GeV}$ and $\mu_t=100\rm{GeV}$ such that:
\begin{align}
{\cal{B}}(\Sigma^+\to p\nu\bar{\nu})\left[{\cal O}(m^0_s) \right]&=(4.86\pm0.04)\times10^{-13},
\quad{\cal{B}}(\Sigma^+\to p\nu\bar{\nu})\left[{\cal O}(m^0_s) + {\cal O}(m^1_s)\right]=(5.01\pm0.08)\times10^{-13}.
\end{align}

Next, we consider the uncertainty of the branching ratio of $\Sigma^+\to p\nu\bar{\nu}$ and $\Lambda\to n\nu\bar{\nu}$ in the case of NLO with the energy scale $\mu_c=1\rm{GeV}$ and $\mu_t=100\rm{GeV}$. This uncertainty comes from the parameters $F=0.463\pm 0.008$ and $D=0.804\pm0.008$ \cite{Yang:2015era} in the form factor $g_1(0)$:
\begin{align}
{\cal{B}}(\Sigma^+\to p\nu\bar{\nu})&= (5.01 \pm 0.12)\times10^{-13},\quad{\cal{B}}(\Lambda\to n\nu\bar{\nu})= (2.03 \pm 0.05)\times10^{-12}.
\end{align}
For the decay $\Omega^-\to \Xi^-\nu\bar{\nu}$,  $C_5^A(0)=1.653\pm0.006$ in the SU(3) symmetry, while $C_5^A(0)=1.612\pm0.007$ in the SU(3) symmetry breaking~\cite{Yang:2015era}. In the case of NLO with the energy scale $\mu_c=1~{\rm GeV}$ and $\mu_t=100~{\rm GeV}$ the  branching ratio ${\cal{B}}(\Omega^-\to \Xi^-\nu\bar{\nu})$ is then calculated as:
\begin{align}
{\cal{B}}(\Omega^-\to \Xi^-\nu\bar{\nu})(sy)&= (1.84 \pm 0.01)\times10^{-11},\quad{\cal{B}}(\Omega^-\to \Xi^-\nu\bar{\nu})(br)=  (1.93 \pm 0.01)\times10^{-11}.
\end{align}

As an illustration of the effects of $q^2$ distribution in the form factors, we attempt to use the following parametrization for all form factors:
\begin{align}
F(q^2)=\frac{F(0)}{1-\frac{q^2}{m^2}},
\end{align}
with $m$ representing the initial hyperon mass. For example,  for the NLO case of $\mu_c=1~{\rm GeV}$ and $\mu_t=100~{\rm GeV}$, we obtain:
\begin{align}
{\cal{B}}(\Lambda\rightarrow n\nu\bar{\nu})(F(0))&= 1.98\times10^{-12},\quad{\cal{B}}(\Lambda\rightarrow n\nu\bar{\nu})(F(q^2))=  2.03 \times10^{-12},\nonumber\\
{\cal{B}}(\Sigma^{+}\rightarrow p\nu\bar{\nu})(F(0))&= 5.05\times10^{-13},\quad{\cal{B}}(\Sigma^{+}\rightarrow p\nu\bar{\nu})(F(q^2))=  5.16 \times10^{-13}.
\end{align}
We find that the differences between the two cases are small, about a  few percent.

When all the above errors in the form factors are included, we find that the final branching ratios for most decay modes have an uncertainty of about $5\%$ to $10\%$.

\subsection{Contribution from MSSM}
The effective Hamiltonian for $s\to d\nu\bar{\nu}$ in the generalized supersymmetry (SUSY) extension of SM is given in
Eq. (\ref{eq:Heff_SM}), with $X(x_{t})$ replaced by \cite{Buras:1997ij}
\begin{equation}
X_{{\rm new}}=X(x_{t})+X_{H}(x_{tH})+C_{\chi}+C_{N}.
\end{equation}
Here, $x_{tH}=m_{t}^{2}/m_{H^{\pm}}^{2}$, and $X_{H}(x_{tH})$ corresponds
to the charged Higgs contribution. $C_{\chi}$ and $C_{N}$ denote
the chargino and neutralino contributions
\begin{align*}
C_{\chi} & =X_{\chi}^{0}+X_{\chi}^{LL}R_{s_{L}d_{L}}^{U}+X_{\chi}^{LR}R_{s_{L}t_{R}}^{U}+X_{\chi}^{LR*}R_{t_{R}d_{L}}^{U},\\
C_{N} & =X_{N}R_{s_{L}d_{L}}^{D},
\end{align*}
where $X_{\chi}^{i}$ and $X_{N}$ depend on the SUSY masses, and respectively
on the chargino and neutralino mixing angles. The explicit expressions
for $X_{H}(x)$, $C_{\chi}$ and $C_{N}$ can be found in Ref. \cite{Buras:1997ij}.
The $R$ parameters are defined in terms of mass insertions, and their upper limits are listed in TAB.~\ref{Tab:R_upper_limit}~\cite{Buras:1997ij}.
It should be mentioned that the phase $\phi$ of $R_{s_{L}t_{R}}^{U}$
and $R_{t_{R}d_{L}}^{U}$ is a free parameter which ranges from 0
to $2\pi$. We set $\phi=0$ as a central result.
\begin{table}
\caption{Upper limits for the $R$ parameters. Notice that the phase of $R_{s_{L}t_{R}}^{U}$ and $R_{t_{R}d_{L}}^{U}$ is unconstrained.}
\label{Tab:R_upper_limit}
\begin{center}
\begin{tabular}{c|c}
	\hline\hline
	quantity & upper limit\tabularnewline
	\hline
	$R_{s_{L}d_{L}}^{D}$ & $(-112-55i)\frac{m_{\tilde{d}_{L}}}{500{\rm GeV}}$\tabularnewline
	\hline
	$R_{s_{L}d_{L}}^{U}$ & $(-112-54i)\frac{m_{\tilde{u}_{L}}}{500{\rm GeV}}$\tabularnewline
	\hline
	$R_{s_{L}t_{R}}^{U}$ & ${\rm Min}\{231\left(\frac{m_{\tilde{u}_{L}}}{500{\rm GeV}}\right)^{3},43\}\times e^{i\phi},0<\phi<2\pi$\tabularnewline
	\hline
	$R_{t_{R}d_{L}}^{U}$ & $37\left(\frac{m_{\tilde{u}_{L}}}{500{\rm GeV}}\right)^{2}\times e^{i\phi},0<\phi<2\pi$\tabularnewline
	\hline\hline
\end{tabular}
\end{center}
\end{table}

The parameters in TAB.~\ref{tab:constr} are adopted for detailed calculations~\cite{Buras:2004qb}. The assumption $M_{1}\approx 0.5M_{2}$ was made~\cite{Martin:1997ns}.
\begin{table}[htbp]
\caption{Parameters and their ranges used in the paper\cite{Buras:2004qb}. All mass parameters are in GeV.}
\label{tab:constr}
\begin{center}
\begin{tabular}{c|c|c|c|c}\hline\hline
parameters\cite{Buras:2004qb}
&the meaning of parameters\cite{Buras:2004qb}
&the range of parameters\cite{Buras:2004qb}
&Set.I\cite{Buras:2004qb}
&Set.II\cite{Buras:2004qb}\\\hline
$\beta$
&The angle of unitarity triangle
&$-180^{\circ}\leq \beta \leq 180^{\circ}$
&$\tan \beta=2$
&$\tan\beta=20$\\
$M_A$
&CP-odd Higgs boson mass
& $150\leq M_A \leq 400$
&$333$
&$260$ \\
$M_2$
&$SU(2)$ gaugino mass; we use $M_1$ GUT-related to $M_2$
&
$50\leq M_2 \leq 800$
&$181$
&$750$\\
$\mu$
&Supersymmetric Higgs mixing parameter
 & $-400\leq \mu \leq 400$
&$-375$
&$-344$\\
$M_{sl}$
&Common flavour diagonal slepton mass parameter
&$95\leq M_{sl} \leq 1000$
&$105$
& $884$\\
$M_{sq}$
&Common mass parameter for the first two generations of squarks
& $240\leq M_{sq} \leq 1000$
&$308$
&$608$ \\
$M_{\tilde t_L}$
&Squark mass parameter for the right stop
&$50\leq M_{\tilde t_R} \leq 1000$
&$279$
&$338$\\\hline\hline
\end{tabular}
\end{center}
\end{table}
With the above parameters, the branching ratios of
hyperon decays are listed in TAB.~\ref{Tab:branchingfraction}, and are significantly enhanced compared with the SM results. Taking as examples the decays $\Lambda\to n\nu\bar{\nu}$ and $\Omega^-\to\Xi^{-}\nu\bar{\nu}$ with the energy scale $\mu_c=1~{\rm GeV}$ and $\mu_t=100~{\rm GeV}$, we obtain:
\begin{align}
{\rm NLO} :\quad{\cal B}(\Lambda\to n\nu\bar{\nu})&=1.98\times10^{-12},\quad
{\cal B}(\Omega^-\to\Xi^{-}\nu\bar{\nu})=1.93\times10^{-11},\\
{\rm Set.I} :\quad{\cal B}(\Lambda\to n\nu\bar{\nu})&=8.14\times10^{-12},\quad
{\cal B}(\Omega^-\to\Xi^{-}\nu\bar{\nu})=7.94\times10^{-11},\\
{\rm Set.II} :\quad{\cal B}(\Lambda\to n\nu\bar{\nu})&=3.78\times10^{-12},\quad
{\cal B}(\Omega^-\to\Xi^{-}\nu\bar{\nu})=3.69\times10^{-11}.
\end{align}
Comparing the results
of NLO+SUSY (Set. I) and (Set. II) with the ones of NLO, we see that all the branching ratios are roughly enhanced by a factor of 4 and 2, respectively. However, none of these results can be probed at the ongoing experimental facilities, like BESIII experiment~\cite{Li:2016tlt}.

\subsection{Contribution from minimal 331 model}
The so-called  minimal $331$ model is an extension of SM at the TeV scale, where the weak gauge group of SM $\rm{SU(2)}_L$ is extented to $\rm{SU(3)}_L$. In this model, a new neutral $Z^{\prime}$ gauge boson can give very important additional contributions, for it can transmit FCNC at the tree level. In TAB.~\ref{Tab:branchingfraction}, we denote this model as M331.
More details of this model can be found in Ref. \cite{Ng:1992st}.
The minimal 331 model leads to a new term in the effective Hamiltonian~\cite{Promberger:2007py}:
\begin{equation}
{\cal H}^{Z^{\prime}}_{\rm eff}=\sum_{l=e,\mu,\tau}\frac{G_{F}}{\sqrt{2}}\frac{\tilde V_{32}^{*}\tilde V_{31}}{3}\Big(\frac{M_{Z}}{M_{Z^{'}}}\Big)^{2}\cos^{2}\theta_{W}(\bar{s}d)_{V-A}(\bar{\nu}_{l}\nu_{l})_{V-A}+{\rm h.c.},\label{eq:Heff_Z'}
\end{equation}
with $M_{Z^{\prime}}=1~{\rm TeV}$,
${\rm Re}[(\tilde V_{32}^{*}\tilde V_{31})^2]=9.2\times 10^{-6}$ and ${\rm Im}[(\tilde V_{32}^{*}\tilde V_{31})^2]=4.8\times 10^{-8}$~\cite{Promberger:2007py}. The other parameters are the same as the SM inputs \cite{Tanabashi:2018oca,Buras:2005gr}.
The function $X(x_{t})$ in Eq.~(\ref{eq:Heff_SM}) can be redefined as
 $X(x_{t})=X^{\rm SM}(x_t)+\Delta X$ with
 \begin{equation}
\Delta X=\frac{\sin^2\theta_{W}\cos^2\theta_{W}}{\alpha}\frac{2\pi}{3}\frac{\tilde V_{32}^{*}\tilde V_{31}}{V_{ts}^{*}V_{td}}\Big(\frac{M_{Z}}{M_{Z^{\prime}}}\Big)^{2}.\label{eq:delta_X}
\end{equation}

With the modified function $X(x_{t})$ and considering the NLO contribution, the branching ratios of rare hyperon decays in the minimal 331 model can be calculated, as shown in TAB.~\ref{Tab:branchingfraction}. The NLO+M331 predictions are much larger than the NLO results in SM, and are two and four times larger than the results of  NLO+SUSY (Set. I) and NLO+SUSY (Set. II), respectively.
\section{Conclusions}
FCNC processes offer important tools to test SM, and to search
for possible new physics. The two decays $K^{+}\to\pi^{+}\nu\bar{\nu}$
and $K_{L}\to\pi^{0}\nu\bar{\nu}$ have been widely studied, while
the corresponding baryon sector has not been explored.
In this work, we studied the $s\to d\nu\bar{\nu}$ rare hyperon decays. We adopted the leading order approximations for
the form factors for small $q^{2}$, and derived expressions for the decay
width. We applied the decay width formula to both SM
and new physics contributions. Different energy scales were considered. The
branching ratios in SM range from $10^{-14}$ to $10^{-11}$, and the largest is of the same order as for the decays $K^+\to\pi^{+}\nu\bar{\nu}$ and $K_{L}\to\pi^{0}\nu\bar{\nu}$. When all the errors in the form factors are included, we found that the final branching ratios for most decay modes have an uncertainty of about $5\%$ to $10\%$. After taking into account the contribution from MSSM, the branching ratios are enhanced by a factor of $2 \sim 4$. The branching ratios of hyperon decays in the minimal 331 Model are seven times larger than the SM results.
\section*{Acknowlegement}
The authors are grateful to Profs. Hai-Bo Li and Wei Wang for useful discussions.
This work is supported  in part   by National  Natural
Science Foundation of China under Grant No.11575110, 11735010, 11911530088,  Natural  Science Foundation of Shanghai under Grant No.~15DZ2272100, and  by Key Laboratory for Particle Physics, Astrophysics and Cosmology, Ministry of Education.

\end{document}